\documentclass[12pt, a4paper]{article}

\usepackage[margin=1in]{geometry}
\usepackage{authblk} 
\usepackage{color}
\usepackage{amsmath,amssymb,bm,epsfig,afterpage}
\usepackage{graphicx} 
\usepackage{fancyhdr} 
\usepackage{amsfonts}
\usepackage{graphicx}
\usepackage{tabularx} 
\usepackage[
singlelinecheck=false 
]{caption}
\usepackage{cite}

\def\be{\begin{equation}}
\def\ee{\end{equation}}
\def\bea{\begin{eqnarray}}
\def\eea{\end{eqnarray}}

\def\ba{\begin{array}}
\def\ea{\end{array}}

\newcommand{\ol}[1]{\overline{#1}}
\newcommand{\order}[1]{\mathcal{O}\left({#1}\right)}
\newcommand{\MNS}{\mathrm{PMNS}}
\newcommand{\abs}[1]{\left|{#1}\right|}
\newcommand{\Lcal}{\mathcal{L}}
\newcommand{\eV}{\mathrm{eV}}
\newcommand{\Acal}{\mathcal{A}}
\newcommand{\vev}[1]{\langle{#1}\rangle}
\newcommand{\ndbd}{{0\nu\beta\beta}}
\newcommand{\MeV}{\mathrm{MeV}}
\newcommand{\GeV}{\mathrm{GeV}}
\newcommand{\TeV}{\mathrm{TeV}}
\newcommand{\Dp}{{D^\prime}}
\newcommand{\vHp}{v_{H^\prime}}
\newcommand{\eff}{\mathrm{eff}}
\newcommand{\tnu}{{\tilde{\nu}}}
\newcommand{\eps}{\epsilon}
\newcommand{\wt}[1]{\widetilde{#1}}

\begin{document}

\renewcommand\headrule{} 
\title{\Large \bf{Neutrinos in Global SU(5) F-theory Model}}
\author{Stuart Raby}
\affil{\emph{Department of Physics}\\\emph{The Ohio State University}\\\emph{191 W.~Woodruff Ave, Columbus, OH 43210, USA}}
\author{Junichiro Kawamura} 
\affil{\emph{Center for Theoretical Physics of the Universe, Institute for Basic Science (IBS),
Daejeon 34051, Korea}}

\maketitle
\pagenumbering{gobble} 

\begin{abstract}
\normalsize\parindent 0pt\parskip 5pt
In this talk, given at Corfu 2022 Workshop on the Standard Model and Beyond, I present work in collaboration with Junichiro Kawamura,Ref.~\cite{Kawamura:2022ygc}. The talk is also based on a number of papers on a Global $SU(5)$ F-theory GUT in collaboration with Herb Clemens.  In the model $SU(5)$ is broken to the MSSM via a Wilson line.   This is accomplished (without problems with vector-like exotics) by simultaneously describing the F-theory model and its Heterotic dual.   The model has a twin MSSM sector and it's the neutrino sector of the field I consider in the talk.
\end{abstract}


\clearpage

\pagenumbering{arabic} 


\section{Introduction}

In Ref.~\cite{Clemens:2019dts}, a model with $SU(5)\times SU(5)^\prime \times U(1)_X$ gauge symmetry is realized
utilizing Heterotic-F-theory duality and a $4+1$ split of the F-theory spectral
divisor~\cite{Clemens:2019wgx,Clemens:2019mvs}.\footnote{This construction was shown to solve the theoretical problem of constructing an $SU(5)$ F-theory model with Wilson line breaking.  It was shown that Wilson line breaking, \cite{Donagi:2008ca,Beasley:2008kw}, resulted in massless vector-like exotics.  As a result it was argued that a non-flat hypercharge flux was necessary to solve this problem.  This causes problems with gauge coupling unification. On the flip side, it was known that Wilson line breaking of an $SU(5)$ Heterotic model does not have the same problem.
Therefore by constructing the F-theory model with an explicit Heterotic dual it became clear how to solve
the problem of vector-like exotics with Wilson line breaking.}
The Grand Unified Theory (GUT) surface is invariant under
a $\mathbb{Z}_2$ involution which allows for the gauge symmetry to be broken down
to $G_{\mathrm{SM}}\times G_{\mathrm{SM}^\prime}$,
where $G_{SM}$ and $G_{SM^\prime}$ are the Standard Model (SM) gauge group
and its twin counterpart
with Wilson line symmetry breaking~\footnote{The problem of Wilson line breaking
resulting in massless vector-like exotics, emphasized in Refs. \cite{Donagi:2008ca,Beasley:2008kw}, is resolved in the F-theory model with a bi-section and the $\mathbb{Z}_2$ involution including a translation by the difference of the two sections.}.  To summarize, the Minimal Supersymmetric Standard Model (MSSM)
is realized after $SU(5)$ breaking by a Wilson line.
There are no vector-like exotics and R-parity
 and a $\mathbb{Z}_4^R$ symmetry arise in this construction.
The $SU(5)^\prime$, broken to $G_{\mathrm{SM}^\prime}$
corresponds to the twin sector
whose matter content is the same as that in the MSSM sector,
but with a different value for the GUT scale and the GUT coupling constant~\cite{Clemens:2020grq}.
Models with a twin sector, or sometimes called the mirror sector,
have been described as the parity solution to the strong CP problem in the literature~\cite{PhysRevD.41.1286,Barr:1991qx,Gu:2011yx,Gu:2012in,Gu:2013nya,Abbas:2017hzw,Abbas:2017vle,Gu:2017mkm,Hall:2018let,Kawamura:2018kut,Dunsky:2019api,Berbig:2022hsm}.
The possibility of light sterile neutrinos from the mirror sector
is discussed in Ref.~\cite{Zhang:2013ama}.
In this case there is the mirror sector of the SM without supersymmetry (SUSY).
There are only two right-handed neutrinos,
so that the light sterile neutrino is explained together
with asymmetric dark matter (DM)
from the right-handed neutrino decays~\cite{An:2009vq}.

In this talk, we study the phenomenology of the neutrinos in the F-theory model.
The number of right-handed neutrinos has not been determined in the F-theory model.
So by assumption we assume there are only three generations of right-handed neutrinos
which couple to both the MSSM and twin sectors via Yukawa couplings,
as pointed out in Ref.~\cite{Clemens:2019flx}.\footnote{Right-handed neutrinos are
special in the F-theory model.  The right-handed neutrino curve resides in the base $B_3$, unlike
$SU(5)$ matter curves which reside solely on the GUT surface.   As a result, the right-handed
neutrino curve in the Heterotic limit is identified on the visible and twin sectors.}
The tiny neutrino masses can be explained by the type-I seesaw mechanism.
Three of the neutrinos are massless at the tree-level
since there are only three generations of right-handed neutrinos.
The masses of these states are generated through loop corrections,
and hence their masses are expected to be
$\sim m_D^2/(16\pi^2 M)$,
where $m_D$ is the Dirac neutrino mass which may be at the electroweak (EW) scale
and $M$ is the Majorana mass.
The masses of the other three states are $\sim m_{D^\prime}^2/M$,
where $m_{D^\prime}$ is the Dirac mass for the twin neutrinos
which may be at the twin EW scale.
Thus there are three sterile neutrinos
whose mass are also tiny due to the type-I seesaw mechanism at the tree-level.
We shall study the phenomenology of the active neutrinos
and their mixing with the sterile neutrinos.

\section{The Model}
\subsection{Heterotic side} 
The model is defined in terms of an $E_8 \times E_8$ gauge group on an elliptically fibered CY3
which is a torus fibered over a base $B_2$. $E_8$ is broken to $[SU(5)\times U(1)_X]$ gauge by an
$[SU(4)\times U(1)_X]$ Higgs vector bundle.  

\begin{figure}[h!]
\centering
\includegraphics[width=12cm]{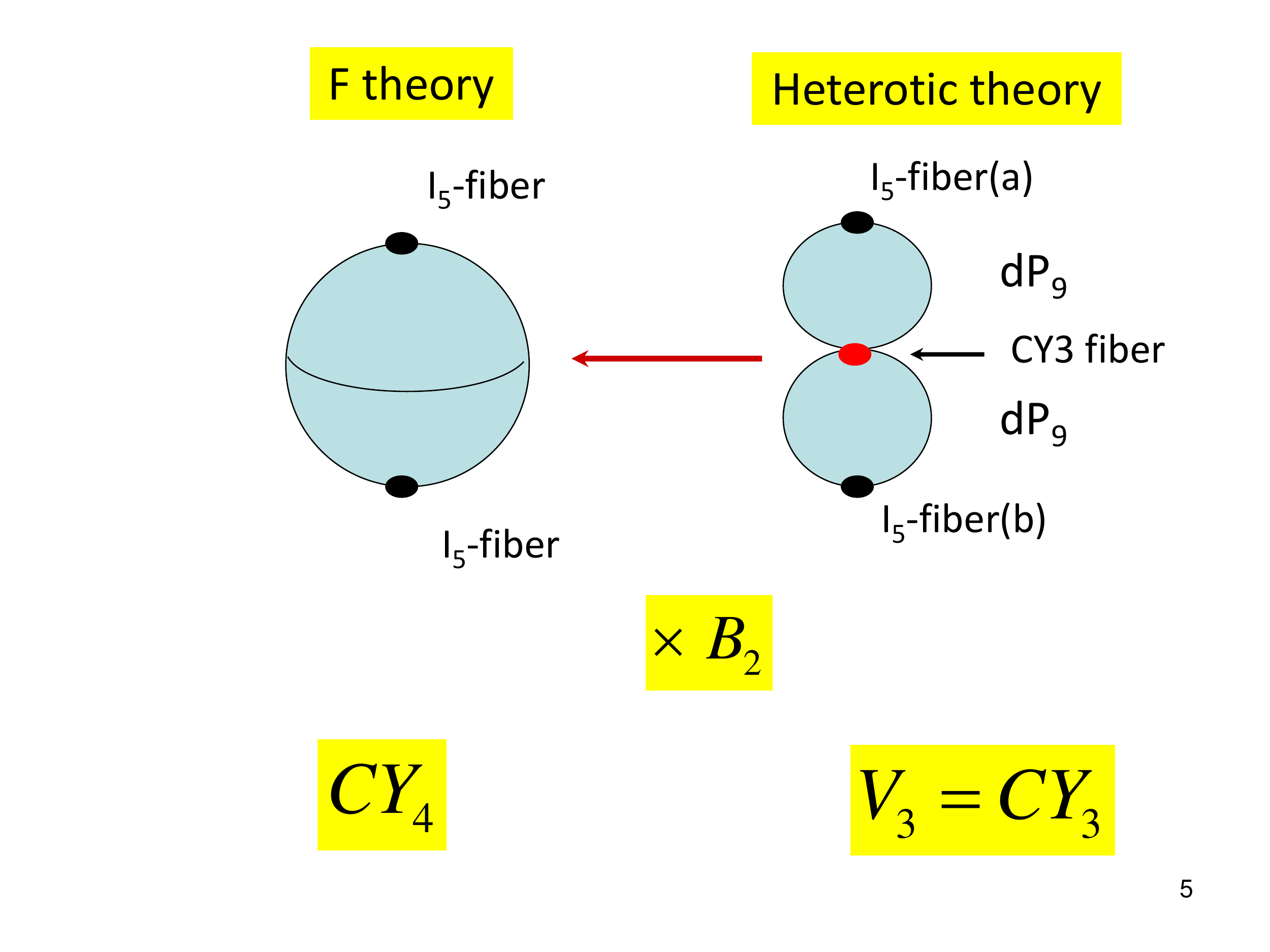}
 \captionsetup{justification=centering}
\caption{Heterotic to F-theory duality}
\label{fig:cy3}
\end{figure}

The CY3 has a freely acting $\mathbb{Z}_2$ involution (preserving the gauge symmetry). The fundamental
group of the manifold downstairs is $\Pi_1(CY3) = \mathbb{Z}_2$.  A hypercharge Wilson line wraps 
the non-contractible cycle and breaks $SU(5)$ gauge to the Standard Model gauge group. The Higgs data is given,
in the semi-stable degeneration limit, in terms of $dP_9 ~U ~dP_9$ connected along an elliptic fiber.  This
defines what is known as the spectral cover.  In Fig. \ref{fig:cy3} the transition from the Heterotic theory to its
F-theory dual is outlined.

\subsection{F-theory side}
We construct a CY4 which is defined by an elliptic fiber over a base $B_3$ with two sections.  It is mathematically defined by a Tate form (or Weierstrass function)
\be w y^2 = x^3 + a_5 w x y + a_4 z w x^2 + a_3 z^2 w^2 y + a_2 z^3 w^2 x + a_0 z^5 w^3 . \ee
The Tate form represents the parameters of the Higgs vector bundle which breaks $E_8 \times E_8$ to a subgroup.
In the Tate form the $[w,x, y]$ are the coordinates of the elliptic fiber.  $z=0$ is the position where
the Weierstrass descriminant vanishes and defines the GUT surface, $S_{GUT}$.   Finally, $z, a_j$ are functions
on $B_3$.
The fiber has two sections.  The first is given by
\be \zeta(b_3) = \{ [ w, x, y] = [0, 0, 1]\}. \ee
We choose the constraint \be a_5 + a_4 +a_3 + a_2 + a_0 = 0. \ee
This gives the second section
\be \tau(b_3) = \{ [ w, x, y] = [1, z^2, z^3]\}. \ee

Now to understand the consequence of this choice,  let $w=1, x = t^2, y= t^3$ and define $s = z/t$.
We then have  \be C \equiv a_5 + a_4 s + a_3 s^2 + a_2 s^3 + a_0 s^5 \ee which defines the
spectral cover.   We can also see that
\be C = (a_5 + a_{54} s - a_{20} s^2 - a_0 s^3 - a_0 s^4)(1 -s) \ee where
$a_{ij} \equiv  a_i + a_j$. This gives a 4 + 1 split of the spectral cover.

\subsection{The Model}

Let me just give some properties of the model without any proof.  For a proof one should look at the references
given in the beginning.   
\begin{itemize}
\item The model explicitly satisfies Heterotic-F-theory duality.  
\item On the Heterotic side there is $E_8 \times E_8$ broken to the gauge group $(SU(5) \times U(1)_X)^2$.  On the F-theory side, as a consequence of the 4 + 1 split of the spectral cover,  the Higgs bundle, $[SU(4)\times U(1)_X]$, lives on a 4 sheeted cover of the GUT surface.    The $U(1)_X$ is a consequence of the 4 + 1 split of the spectral cover.
\item There is a freely acting $\mathbb{Z}_2$ involution acting on the GUT surface.  Downstairs we obtain
$S_{GUT} = Enriques$.
\item The gauge group $SU(5) \times U(1)_X$ is broken to the $SU(3) \times SU(2) \times U(1)_Y \times U(1)_X$ gauge group
with Wilson line breakiing.
\item  There are NO vector-like exotics, since the involution includes a translation by $\zeta(b_3) - \tau(b_3)$.
\item The model has an R-parity and a $\mathbb{Z}_4^R$ symmetry which prevents both dimension 4 and 5 baryon and lepton
number violating operators.
\end{itemize}

{\bf Spectrum} 
\medskip

The model has a visible sector with 3 families of quarks and leptons and one Higgs pair.  The matter curves live on the GUT surface.
The $U(1)_X$ charges of the matter fields are given by the superscripts
\be 10^{-1}_m,  \; \bar 5^{+3}_m,  \; 5^{+2}_h + \bar 5^{-2}_h \ee which allows for Yukawa couplings
\be 10_m \bar 5^m \bar 5^h, \; 10_m 10_m 5_h \;\; \rm{but \; NOT} \;\; 10_m \bar 5_m \bar 5_m . \ee
Downstairs, after the involution, the $z=0$ is a two sheeted cover of the GUT surface.  As a consequence
the model includes a twin sector whose spectrum is identical to the visible sector.

The model contains right-handed [RH] neutrinos, see Ref.  ~\cite{Clemens:2019flx}.
The RH neutrino curve, $\Gamma^{-5} \equiv N^{-5}_m$, lives in $B_3$ along with one other curve,
$\Lambda^{+10} \equiv \Phi^{+10}$.   This allows for the possible Yukawa couplings
\be N^{-5}_m \bar 5^{+3}_m 5^{+2}_h , \ee i.e. Dirac neutrino masses
and \be N^{-5}_m N^{-5}_m \Phi^{+10} . \ee  

Note, under the involution,  $U(1)_X$ is broken to $\mathbb{Z}_2$ matter parity.  Moreover, if
$\Phi^{+10}$ obtains a VEV this gives a Majorana mass to the RH neutrinos.
\medskip

{\bf Relative scales of the visible and twin (hidden) sectors, Ref. ~\cite{Clemens:2020grq}}
\medskip

\begin{figure}[h!]
\centering
\includegraphics[width=12cm]{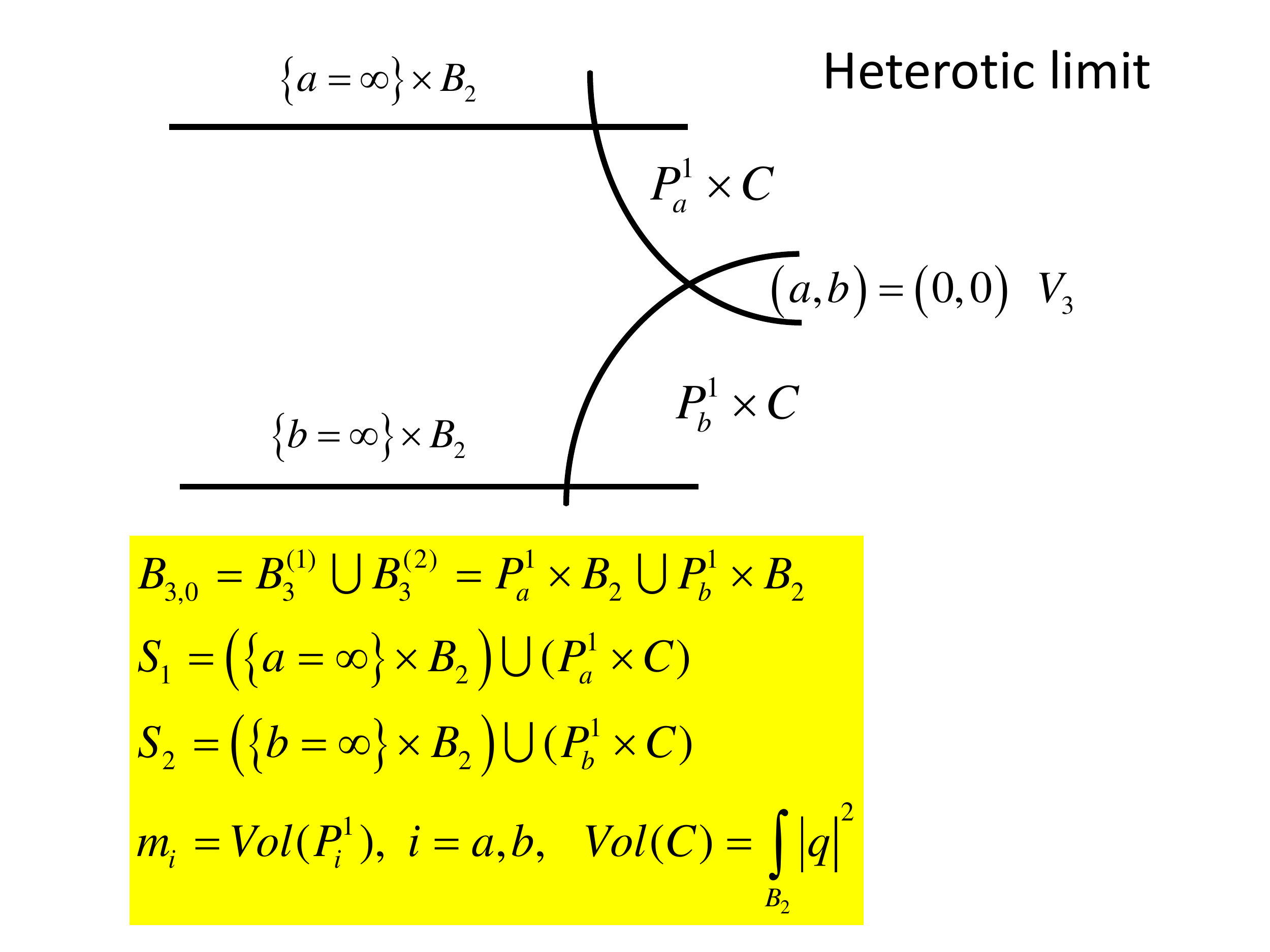}
 \captionsetup{justification=centering}
\caption{In the Heterotic limit the two gauge sectors separate.  The parameter $\delta$ ranges from 0 to 1 where
$\delta = 0$ is the F-theory limit and $\delta = 1$ is the Heterotic limit. The figure describes the $dP_9 \bigcup dP_9$ which intersect on the elliptic curve.  $V_3 = CY3$.}
\label{fig:heteroticlmt}
\end{figure}

In the Heterotic limit of the theory we obtain the gravity action and two gauge actions.  See fig. \ref{fig:heteroticlmt}.
The gravity action is given by \be S_{EH} \sim M_*^8 \int_{\mathbb{R}^{3,1} \times B_3} R \sqrt{-g_\delta} d^{10}x \ee
with $M_{Pl}^2 \simeq M_*^8 \cdot Vol(B_{3,\delta})$.  While the gauge action is given by
\be S_{gauge} \sim - M_*^4 \int_{\mathbb{R}_{3,1}\times S_i} \left( Tr(F_1^2) \sqrt{-g_1} + Tr(F_2^2) \sqrt{-g_2}\right) \delta^2(z_0) d^{10}x .\ee  We have two gauge groups in the Heterotic limit.

We have \be \alpha_G^{-1}(i) \sim M_*^4 Vol(S_i) \ee and \be M_G(i)^{-4} \sim Vol(S_i) \ee where 
\be Vol(S_i) = Vol(B_2) + m_i Vol(C),  \;\; m_i = Vol(P^1_i), \; i = a,b \equiv  1,2. \ee
Thus \be \alpha_G(i) M_{Pl} =  \sqrt{\frac{(m_1 + m_2) Vol(B_2)}{Vol(B_2)(1 + K m_i)}}.\ee
We have \be \alpha_G(2)/\alpha_G(1) = \frac{1+K m_1}{1 + K m_2} , \;\; M_G(2)/M_G(1) =  \left(\frac{1+K m_1}{1 + K m_2} \right)^{1/4} .\ee
For example if we take for the visible sector $\alpha_G(1)^{-1} = 24, \, M_G(1) = 3 \times 10^{16}$ GeV and
for the twin sector, take $\alpha_G(2)^{-1} = 8.7, \, M_G(1) = 3.9 \times 10^{16}$ GeV we need
$\frac{1+K m_1}{1 + K m_2} = 2.8$.
\bigskip

{\bf Summary}
\medskip

\begin{itemize}
\item We constructed a Global $SU(5)$ F-theory model with Wilson line breaking.
\item The Wilson line wraps the GUT surface breaking $SU(5)$ to the SM gauge group.
\item It has a complete twin sector with different scales determined by the sizes of the visible and twin manifolds.
\item $M_{GUT} = M_{compactification} \sim 1/R_{cycle}$
\item Non-local GUT breaking by the Wilson line gives precise gauge coupling unification.
\item It contains 3 families and one pair of Higgs doublets and NO vector-like exotics !
\item It has a $\mathbb{Z}_2$ matter parity and a $\mathbb{Z}_4^R$ symmetry.
\item Allowed Yukawa couplings are consistent with what is needed for giving quarks and charged lepton masses and
a See-Saw mechanism for neutrino masses.
\item  Twin matter contains a dark matter candidate ?
\end{itemize}

\subsection{Neutrino Portal}
\medskip

{\bf The Low Energy Theory}
\medskip

We have the MSSM and a twin MSSM$^\prime$.   We assume  $\Lambda_{QCD^\prime} > \Lambda_{QCD}$ which
implies heavier twin baryons. In both the visible and twin sectors there are two pairs of Higgs doublets.
We assume the VEVs in the twin sector
\be \langle H_u^\prime \rangle = \left( \begin{array}{c} 0 \\ v_{H_u^\prime} \end{array} \right) ,  \;\;\;\;
\langle H_d^\prime \rangle = \left( \begin{array}{c} v_{H_d^\prime} \\ 0\end{array} \right) .\ee
We also assume, generically,  $v_{H^\prime} > v_H$ which implies heavier twin Dirac quark and lepton masses.
RH neutrino masses are given by \be y_{ij} \langle \Phi \rangle \Gamma_i \Gamma_j \Rightarrow M_{ij} N_i N_j , \;\; i, j = 1, .., , 3 . \ee  They are identified in the visible and twin sectors with one global $U(1)_X$.

The superpotential is given by 
\begin{align}
 W = \frac{1}{2} N_i M_{ij} N_j
     + N_i Y_{ij} \ell_j H_u + N_i Y_{ij}^\prime \ell^\prime_j H^\prime_u,
\end{align}
where $\ell_i$ and $H_u$ ($\ell_i^\prime$ and $H_u^\prime$)
are the MSSM (twin) lepton and up-type Higgs doublets.

After integrating out the heavy neutrinos (assuming $M \gg v_H^\prime > v_H$) we have
\begin{align}
 W = - \frac{1}{2} (\ell Y^T H_u + \ell^\prime Y^{\prime T} H_u^\prime) M^{-1} (Y \ell H_u + Y^\prime \ell^\prime H^\prime_u) .
\end{align}
Clearly there are 3 massive and 3 massless neutrinos.
\medskip

{\bf Radiative neutrino masses, Ref. ~\cite{Dedes:2007ef}}
\medskip

\begin{figure}[h!]
\centering
\includegraphics[width=12cm]{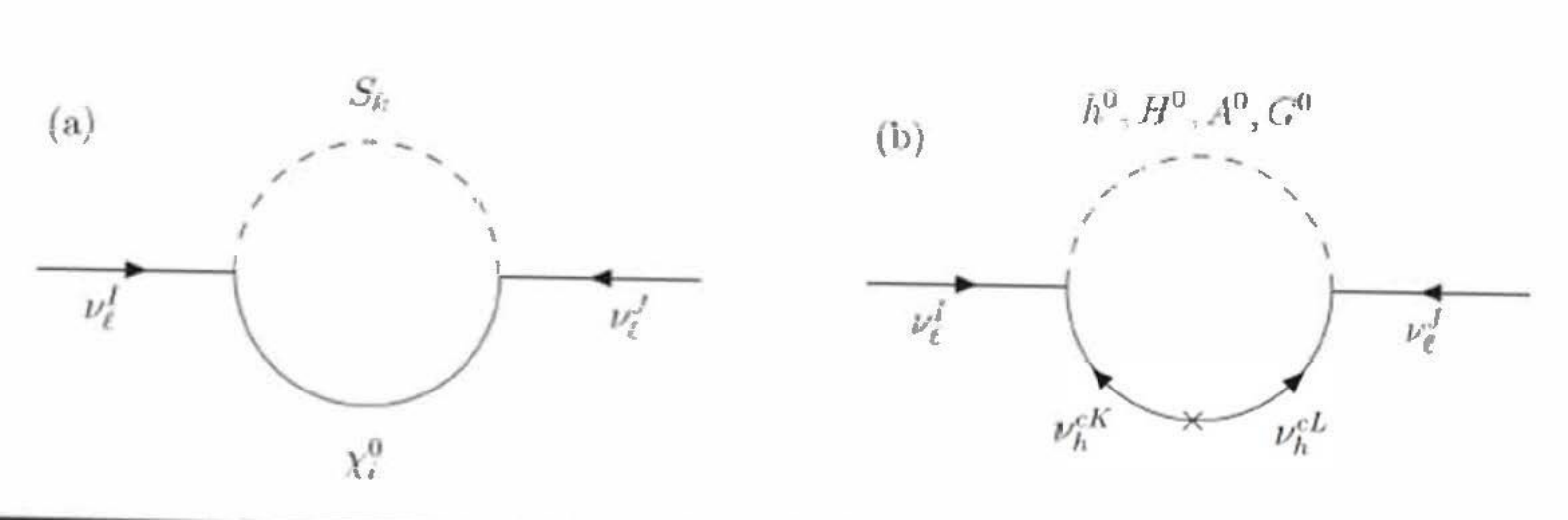}
 \captionsetup{justification=centering}
\caption{In these one loop diagrams, (a)  $S_k (k=1, ..., 6)$ are the light sneutrinos and $\chi_i^0$ are neutralinos. 
(b) These include Higgs, goldstinos and heavy neutrinos.}
\label{fig:radcorr}
\end{figure}

The light neutrino masses, including radiative corrections, are given by
\be M_{\nu_\ell} = \left(  \begin{pmatrix}
 \delta m_{LL} - m_D^T M^{-1} m_D & - m_D^T M^{-1} m_{D^\prime} \\
- m_{D^\prime}^T M^{-1} m_D &  \delta m_{L^\prime L^\prime} - m_{D^\prime}^T M^{-1} m_{D^\prime}
\end{pmatrix} \right) \ee and in the limit $m_{D^\prime} \gg m_D$ we have
\be M_{\nu_\ell} \sim  \left(  \begin{pmatrix}
 \delta m_{LL} & 0 \\
0 &  - m_{D^\prime}^T M^{-1} m_{D^\prime}
\end{pmatrix} \right) .\ee

We define the diagonalization unitary matrix $U_\ell$
for the light neutrino mass matrix with radiative corrections as
\begin{align}
 U_\ell^T M_{\nu_\ell} U_\ell = \mathrm{diag}\left(
m_{\nu_1},m_{\nu_2},m_{\nu_3},m_{\nu_4},m_{\nu_5},m_{\nu_6}
\right),
\end{align}

We decompose $U_\ell$ and parametrize the upper $3\times 6$ block of it as
\begin{align}
U_\ell  =:
\begin{pmatrix}
 A_{\ell} \\ A_{\ell^\prime}
\end{pmatrix},
\quad
A_\ell =:
\begin{pmatrix}
 U_{e1} &  U_{e2} &  U_{e3} &  U_{e4} &  U_{e5} &  U_{e6} \\
 U_{\mu1} &  U_{\mu2} &  U_{\mu3} &  U_{\mu4} &  U_{\mu5} &  U_{\mu6} \\
 U_{\tau1} &  U_{\tau2} &  U_{\tau3} &  U_{\tau4} &  U_{\tau5} &  U_{\tau6} \\
\end{pmatrix},
\end{align}
where the left (right) three columns of $A_\ell$ are
for the active (sterile) neutrinos.

The $W$ boson coupling is given by
\begin{align}
 \Lcal_W
= \frac{g}{\sqrt{2}} W_\mu^- \ol{\psi}_e
           \gamma^\mu P_L A_\ell \Psi_\ell + h.c.,
\end{align}
where $\Psi_\ell$ contains the neutrinos in the mass basis
without the heaviest three states with $\order{M}$ masses.
Here, we choose the flavor basis that the charged lepton Yukawa matrix
is positive diagonal in the gauge basis.
We assume that the Pontecorvo-Maki-Nakagawa-Sakata (PMNS) matrix
for the active neutrinos are almost unitary,
so that the angles in the standard parametrization,
\begin{align}
U_\MNS=
\begin{pmatrix}
1 & 0 & 0 \\
0 & c_{23} & s_{23} \\
0 & -s_{23} & c_{23} \\
\end{pmatrix}
\begin{pmatrix}
c_{13} & 0 & s_{13} e^{-i \delta} \\
0 & 1 & 0 \\
-s_{13} e^{i \delta} & 0 & c_{13} \\
\end{pmatrix}
\begin{pmatrix}
 c_{12} & s_{12}& 0 \\
 -s_{12} & c_{12} & 0 \\
0 & 0 & 1
\end{pmatrix},
\end{align}
are related to the elements in $A_\ell$ as
\begin{align}
s_{12} = \frac{\abs{U_{e2}}}{\sqrt{\abs{U_{e1}}^2 + \abs{U_{e2}}^2 }}
\quad
s_{23} = \frac{\abs{U_{\mu 3}}}{\sqrt{\abs{U_{\mu 3}}^2 + \abs{U_{\tau 3}}^2 }}
\quad
 s_{13} = \abs{U_{e3}},
\end{align}
and
\begin{align}
 e^{i\delta}
= \abs{\frac{U_{e2} U_{e3} U_{\mu 3}}{U_{e1}U_{\tau 3}}}
 \left( 1+
\frac{\sqrt{\abs{U_{e1}}^2 + \abs{U_{e2}}^2}\sqrt{\abs{U_{\tau 3}}^2 + \abs{U_{\mu 3}}^2}}
     {\abs{U_{\mu 3}}^2} \frac{U_{\mu 2}^* U_{\mu 3}}{U_{e2}^* U_{e3}}\right).
\end{align}
Here, we consider the fitted data
with the normal ordering (NO)~\cite{deSalas:2017kay,ParticleDataGroup:2022pth}:
\begin{align}
\Delta m_{12}^2 = (7.55 \pm 0.20)\times 10^{-5}~\eV,
\quad
\Delta m_{23}^2 = (2.424 \pm 0.030)\times 10^{-3}~\eV,
\end{align}
\begin{align}
 s_{12}^2= 0.32\pm 0.02,
\quad
 s_{23}^2= 0.547\pm 0.03,
\quad
 s_{13}^2= 0.0216 \pm 0.0083,
\quad
\delta_\mathrm{CP} = 218\pm38~\mathrm{deg},
\end{align}
and with the inverted ordering (IO)~\cite{deSalas:2017kay,ParticleDataGroup:2022pth}:
\begin{align}
\Delta m_{12}^2 = (7.55 \pm 0.20)\times 10^{-5}~\eV,
\quad
\Delta m_{23}^2 = (-2.50 \pm 0.040)\times 10^{-3}~\eV,
\end{align}
\begin{align}
 s_{12}^2= 0.32\pm 0.02,
\quad
 s_{23}^2= 0.5551 \pm 0.03,
\quad
 s_{13}^2= 0.0220 \pm 0.0076,
\quad
\delta_\mathrm{CP} = 281 \pm 27~\mathrm{deg},
\end{align}
where $\Delta m_{ij}^2 := m_{\nu_j}^2 - m_{\nu_i}^2$.
For reference, typical values of the absolute values of the PMNS matrix is
\begin{align}
\label{eq-absPMNS}
 \abs{U_\MNS} \sim
\begin{pmatrix}
 0.82 & 0.55 & 0.15 \\
 0.31 & 0.60 & 0.74 \\
 0.48 & 0.58 & 0.66
\end{pmatrix}.
\end{align}
In our notation, the lightest neutrino is $\nu_1$ ($\nu_3$)
for the NO (IO) case,
but the sterile neutrinos are ordered by their masses,
so the lightest sterile neutrino is always $\nu_4$.

The Majorana neutrinos can induce neutrino-less double
$\beta$ ($\ndbd$) decay.
The $\ndbd$ decay half-life is given by~\cite{Faessler:2014kka},
\begin{align}
 \left[T^{0\nu}_{1/2} \right]^{-1}
= \Acal\abs{m_p\sum_{i=1}^6 U^2_{ei} \frac{m_{\nu_i}}{\vev{p^2}+m_{\nu_i}^2}}^2,
\end{align}
where the values of $\Acal$ and $\vev{p^2}\simeq (200~\MeV)^2$ are tabulated
in Table.1 of Ref.~\cite{Faessler:2014kka}.
We choose the values which provide the most conservative limits,
i.e. (a) Argonne potential with $g_A=1.00$,
for $\vev{p^2}$ and $\Acal$;
$\sqrt{\vev{p^2}} (^{76}\mathrm{Ge})= 0.159~\GeV$,
$\Acal(^{76}\mathrm{Ge})  = 2.55\times 10^{-10}\;\mathrm{yrs}^{-1}$,
and
$\sqrt{\vev{p^2}} (^{136}\mathrm{Xe})= 0.178~\GeV$,
$\Acal(^{136}\mathrm{Xe})  = 4.41\times 10^{-10}\;\mathrm{yrs}^{-1}$.
The limits for $^{76}$Ge and $^{136}$Xe
are~\cite{KamLAND-Zen:2012mmx,GERDA:2013vls}
\begin{align}
 T_{1/2}^{0\nu}(^{76}\mathrm{Ge})
 \ge 3.0\times 10^{25}~\mathrm{years},
\quad
 T_{1/2}^{0\nu}(^{136}\mathrm{Xe})
 \ge 3.4\times 10^{25}~\mathrm{years},
\end{align}
respectively.
\medskip

\subsection{Simplified analysis}
\medskip

For simplicy we assume $Y^\prime = Y, \;\;  v_H < v^\prime_{H},$  the Majorana mass, $M$ and the Yukawa matrices, $Y^\prime_e = Y_e$, are diagonal with positive elements. We consider $m_H, m_A \gg m_h$ i.e. the decoupling limit and
$\tan\beta^\prime \neq \tan\beta, \;\; M_N = 10^{12}$ ~GeV.   We consider the CMSSM scenario for soft parameters
with $\tan\beta = 10, \;\; \sin\mu = +1, \;\; m_0 = - A_0 = 5$ TeV. and $M_{1/2} = 2.5$ TeV.
We calculate the parameters at the TeV scale
by using \texttt{softsusy-4.1.12}~\cite{Allanach:2001kg}.
At this point, the SM-like Higgs mass is $125.69~\GeV$.
The soft parameters relevant to the neutrino masses are given by
\begin{align}
 M_1 = 1.134~\TeV,
\quad
 M_2 = 2.017~\TeV,
\quad
\mu = 3.424~\TeV, \\
\quad
m_L = (5.2269, 5.2268, 5.1983)~\TeV.
\end{align}
The lightest neutrino, $\nu_1$ for NO and $\nu_3$ for IO,
mass is assumed to be $0.001~\eV$.
We scan over the value of $v_{H^\prime}/v_H$.
We always found the values
which explain the neutrino mixing parameters throughout our scan,
up to numerical errors.

\begin{figure}[h!]
\centering
\includegraphics[width=16cm]{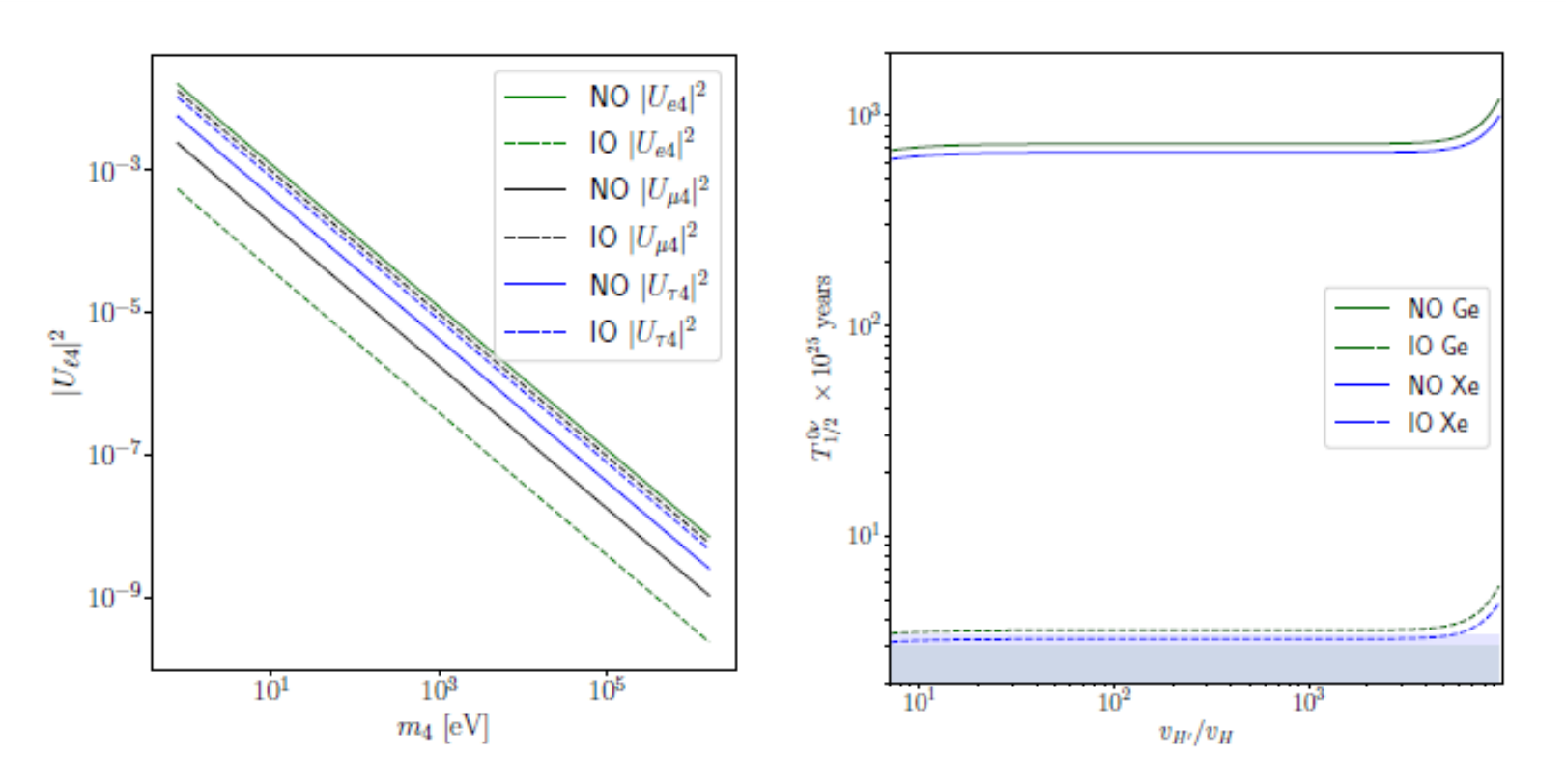}
 \captionsetup{justification=centering}
\caption{The mass and mixing of the lightest sterile neutrino $\nu_4$ (left) and the $0\nu\beta\beta$ decay
with varying $v_{H^\prime}/v_H$ (right).  On the left panel, the solid (dashed) lines correspond to the NO (IO).
The colors are the charged lepton flavors. On the right panel,
the green (blue) lines are the values for Ge (Xe).
The colored regions are the lower bounds on the lifetime. }
\label{fig:neumass}
\end{figure}

The left panel of Fig.~\ref{fig:neumass} shows
the mass and mixing of the lightest sterile neutrino $\nu_4$.
The solid (dashed) lines are the cases of NO (IO) of the active neutrinos.
Since we assume $m_D \propto m_\Dp$,
the mixing matrix for the sterile neutrinos are similar
to the active ones up to the $\order{1}$ coefficients from the soft parameters,
i.e.
\begin{align}
 \begin{pmatrix}
 \abs{ U_{e4}} \\ \abs{U_{\mu 4}} \\ \abs{U_{\tau 4}}
 \end{pmatrix}
\propto
  \begin{pmatrix}
 \abs{ U_{e1}} \\  \abs{ U_{\mu 1}} \\  \abs{ U_{\tau 1}}
 \end{pmatrix}
~\mathrm{for~NO\quad and}~
 \begin{pmatrix}
 \abs{ U_{e4}} \\ \abs{U_{\mu 4}} \\ \abs{U_{\tau 4}}
 \end{pmatrix}
\propto
  \begin{pmatrix}
 \abs{ U_{e3}} \\  \abs{ U_{\mu 3}} \\  \abs{ U_{\tau 3}}
 \end{pmatrix}
~\mathrm{for~IO}.
\end{align}
Thus, $\abs{U_{e4}}$ ($\abs{U_{\mu 4}}$)
is the largest element in the NO (IO) case.

The right panel of Fig.~\ref{fig:neumass} shows the lifetime of $\ndbd$ decays.
Since the sterile neutrinos are much lighter than $\order{100~\MeV}$
for $v_{H^\prime}/v_H \lesssim 10^4$,
the contributions are proportional to $U_{ei}^2 m_{\nu_i}$.
In the NO case, the contributions from the heavier sterile neutrinos
are more suppressed by the mixing angles, see Eq.~\eqref{eq-absPMNS}.
While in the IO case, the heavier states have degenerate masses,
$m_{\nu_5} \simeq m_{\nu_6}$ and the mixing angles are not suppressed.
Therefore the lifetimes are much shorter for the IO case,
and hence $m_{\nu_1} \lesssim 0.001~\eV$ is required to be consistent
with the current limits.

Table~\ref{tab-bench} shows the benchmark points in the NO and IO cases.
The size of $\vHp/v_H$ is chosen
such that the anomaly in the reactor experiments, discussed in the next section,
are explained in the NO case.
We see that the neutrino mixing data is consistent
with the neutrino mixing observables.
The mixing angles involving the sterile neutrinos are much smaller than
those in the active neutrinos, and hence the $3\times 3$ PMNS matrix
is almost unitary.
Since we assume the flavor structure of the Dirac matrices are the same,
the relative sizes of the masses and mixing are similar
among the active and sterile neutrinos.
The lifetime of the eV sterile neutrinos are longer than
$10^{35}~\mathrm{sec}$~\cite{Lee:1977tib,Pal:1981rm,Barger:1995ty,Drewes:2016upu},
so the sterile neutrinos are stable as compared to the age of the universe.

\begin{table}[t]
 \centering
\caption{\label{tab-bench}
The benchmark points in the CMSSM scenario.}
\small
\begin{tabular}[t]{c|cc} \hline
 & NO & IO \\ \hline\hline
$\log_{10} v_{H^\prime}/v_H$ & 0.89 & 0.89 \\
$(d_1, d_2, d_3)~[\mathrm{eV}]$ & (0.0185, -0.1621, 0.9270) & (-0.9129, 0.9271, 0.0185) \\
$(s_{12}^n, s_{23}^n, s_{13}^n, \delta_n)$ & (0.3680, 0.7051, 0.5073, 0.3436) & (0.4601, 0.7575, 0.4231, 0.5122) \\ \hline
$(\Delta m_{12}^2\times 10^5, \Delta m_{23}^2\times 10^{3})~[\mathrm{eV}^2]$ & (7.550, 2.424) & (7.550, -2.500) \\
 $(s_{12}^2, s_{23}^2, s_{13}^2, \delta)$ & (0.320, 0.547, 0.022, -2.478) & (0.320, 0.551, 0.022, -1.379) \\
 $(m_{\nu_4}, m_{\nu_5}, m_{\nu_6})$ [eV] & (1.136, 9.932, 56.784) & (1.136, 55.923, 56.789) \\ \hline
 $\left|\begin{pmatrix} U_{e4} & U_{e5} & U_{e6} \\ U_{\mu4} & U_{\mu5} & U_{\mu6} \\ U_{\tau4} & U_{\tau5} & U_{\tau6} \end{pmatrix}\right|$ & $
 \begin{pmatrix}0.1041 & 0.0714 & 0.0189 \\
0.0401 & 0.0772 & 0.0934 \\
0.0620 & 0.0723 & 0.0850 \\
\end{pmatrix}  $& $
 \begin{pmatrix}0.0191 & 0.1043 & 0.0711 \\
0.0937 & 0.0517 & 0.0696 \\
0.0846 & 0.0524 & 0.0800 \\
\end{pmatrix}$ \\ \hline
$(T^{0\nu}_{1/2}(^{76}\mathrm{Ge}), T^{0\nu}_{1/2}(^{136}\mathrm{Xe}))$ years &$(689.99, 626.67)\times 10^{25}$ & $(3.46, 3.14)\times10^{25}$ \\ \hline
  \end{tabular}
\end{table}

\subsection{Sterile neutrino phenomenology}

\begin{figure}[t]
 \centering
\includegraphics[height=0.6\hsize]{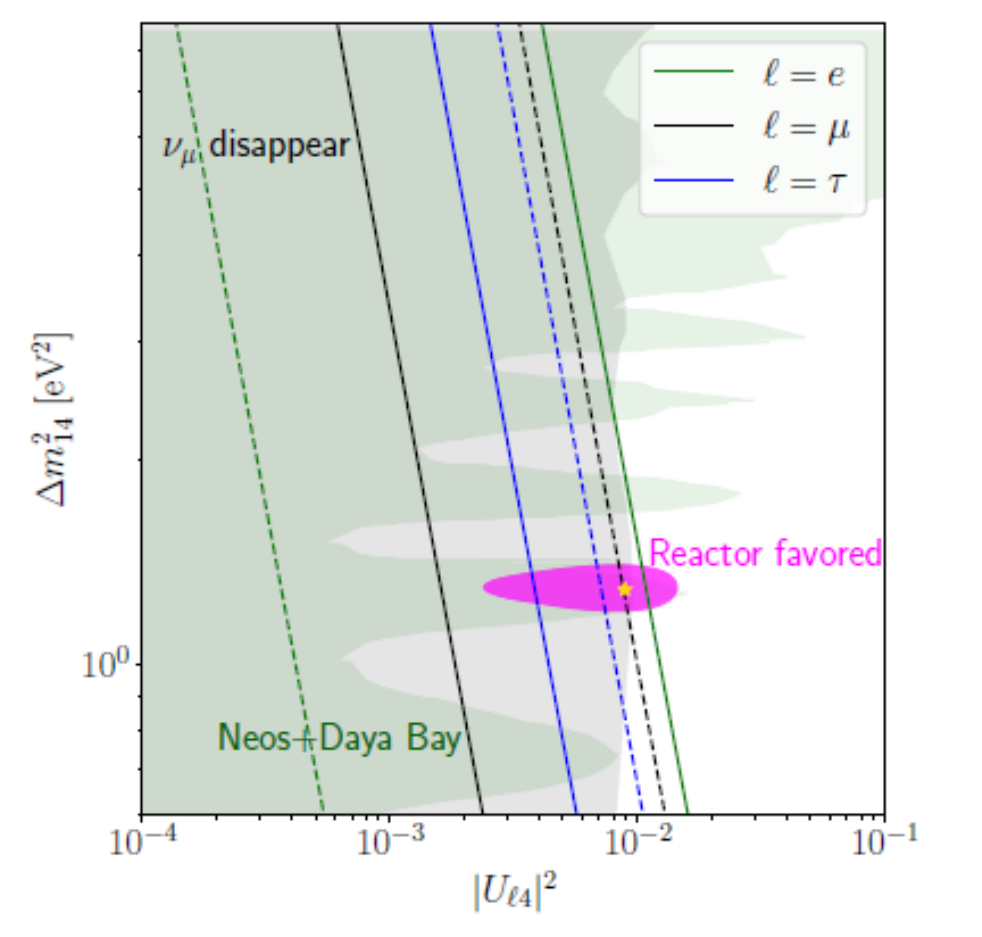}
\caption{\label{fig:mixing}
The mixing angle vs $\Delta m_{14}^2 := m_4^2-m_1^2$.
The green, black and blue lines are $\ell = e, \mu$ and $\tau$, respectively.
The solid (dashed) lines are normal (inverse) hierarchy.
The green (gray) region is allowed by the Neos$+$Daya ($\nu_\mu$ disappearance).
The combined $\nu_e$ disappearance result favors the magenta region.
The yellow star is the best-fit point in the analysis of Ref.~\cite{Dentler:2018sju}.
}
\end{figure}

We study the phenomenology of neutrino mixing
with the lightest sterile neutrino $\nu_4$,
see Refs.~\cite{Boser:2019rta,Dasgupta:2021ies}
for recent reviews of sterile neutrinos.
Under the assumption of $m_\Dp \propto m_D$,
the heavier state $\nu_5$ is about 8~(50) times heavier than $\nu_4$
in the NO (IO), and hence the mixing with these will be sub-dominant~\footnote{
See Refs.~\cite{Giunti:2011gz,Hollander:2014iha}
for the analysis with more than two sterile neutrinos.}.
The following combinations of the mixing with $\nu_4$
are constrained from the reactor experiments~\cite{Dentler:2018sju},
\begin{itemize}
 \item {$\nu_e$ disappearance $\propto |U_{e4}|^2$~\cite{DayaBay:2016qvc,NEOS:2016wee,Alekseev:2016llm}}
 \item {$\nu_\mu\to\nu_e$ oscillations at short baseline $\propto |U_{e4} U_{\mu4}|^2$~\cite{LSND:2001aii,MiniBooNE:2013uba,KARMEN:2002zcm,NOMAD:2003mqg,Antonello:2012pq,OPERA:2013wvp}}
 \item {$\nu_\mu$ disappearance, $\propto |U_{\mu4} |^2$~\cite{IceCube:2014flw,IceCube:2016rnb,MINOS:2017cae,NOvA:2017geg,Super-Kamiokande:2010orq,MiniBooNE:2009ozf,MiniBooNE:2012meu,Wendell:2014dka}}
\end{itemize}
The limit for $\abs{U_{\tau 4}}^2 < 0.13$~\cite{Dentler:2018sju}
is much weaker than limits
on the above combinations~\cite{MINOS:2017cae,NOvA:2017geg}.
In the reactor experiments measuring the $\nu_e$ disappearance,
DayaBay and NEOS put upper bounds on $\abs{U_{e4}}^2$~\cite{DayaBay:2016qvc,NEOS:2016wee},
but the DANSS reported an excess~\cite{Alekseev:2016llm}.
It is interesting that the excess can be explained
consistently with the limits from DayaBay+NEOS,
where $\Delta m_{14}^2 = 1.29~\eV^2$ and $\abs{U_{e4}}^2= 0.0089$~\cite{Dentler:2018sju}.
Anomalies are found in the short base-line experiments
LSND~\cite{LSND:2001aii} and MiniBooNE~\cite{MiniBooNE:2013uba},
which favor $\Delta m_{14}^2 \sim 0.5~\eV^2$
and $4|U_{e4} U_{\mu4}|^2 \sim 0.007$.
The $\nu_\mu$ measurements, however, exclude $|U_{\mu 4}|^2 \lesssim 0.01$
for $\Delta m_{14}^2 \sim \order{0.1-10~\eV^2}$,
and therefore the explanation of the short base-line anomalies
by the mixing with a sterile neutrino
is excluded by the $\nu_\mu$ disappearance result~\cite{Dentler:2018sju}.
Hence we do not consider the anomalies in the short base-line experiments.

Figure~\ref{fig:mixing} shows the favored regions
by the experiments searching for the mixing with a sterile neutrino.
The green (gray) region is allowed by the Neos+Daya Bay
($\nu_\mu$ disappearance) result,
which should be compared with the green (black) lines.
The pink region is the favored region from all the reactor data,
including the DANSS result which observed the anomaly.
In the NO case, the green solid line overlaps the pink region,
and the black line is inside the gray region.
Thus, the anomaly in the DANSS experiment can be explained in this case.
The benchmark point for the NO case in Table~\ref{tab-bench} is chosen
from the overlapped region.
In the IO case, however, the mixing with electron $\abs{U_{e4}}$
is much smaller than the value preferred by the reactor data
for $\Delta m_{14}^2 \sim 1~\eV$.
Furthermore, this case will be excluded by the $\nu_\mu$ disappearance result
even if $\abs{U_{e4}}$ has a certain value
because $\abs{U_{e4}} < \abs{U_{\mu4}}$.
Therefore, the reactor data is fully explained only in the NO case.

\subsection{Cosmology}

The $\eV$ sterile neutrinos which can explain the reactor anomaly may,
however, be incompatible
with cosmological observations~\cite{Dasgupta:2021ies}.
The Planck collaboration obtained the $95\%$ C.L. upper limits
on the effective number of neutrinos $N_\eff < 3.29$
and the sum of the neutrino masses $\sum m_\nu < 0.65~\eV$~\cite{Planck:2018vyg}.
This may have several different types of solutions which I will not get into here.

Before closing, we briefly discuss the cosmology of the other particles in the twin sector.
The twin photon may contribute to $N_\eff$ along with the light sterile neutrinos.
The contribution could be suppressed
if the temperature of the thermal bath of the twin sector
is significantly smaller than the MSSM one.
This requires that the reheating process occurs predominantly in the MSSM sector.
Another possibility is that the twin photon is massive
due to the non-zero VEV of the charged Higgs or sparticles.
This would be the case, for instance,
if anomaly mediation~\cite{Randall:1998uk,Giudice:1998xp}
is the dominant source for SUSY breaking in the twin sector.
In our model, the photon to twin photon kinetic mixing,
$\eps F^{\mu\nu} F^\prime_{\mu\nu}$,
where $F_{\mu\nu}$ ($F^\prime_{\mu\nu}$) is the field strength of the (twin) photon,
is expected to be tiny.
There is a 3-loop diagram which is mediated by neutrinos
whose order is estimated as
\begin{align}
 \eps \sim \frac{g^3g^{\prime 3} m_{\nu^\prime}^2}{(16\pi^2)^3 m_W m_{W^\prime} }
      \sim 10^{-25}\times
          \left(\frac{ \sqrt{gg^\prime}}{0.5}\right)^6
           \left(\frac{m_{\nu^\prime}}{50~\eV}\right)^2
           \frac{m_{W}}{m_{W^\prime}}.
\end{align}
Therefore it is negligibly small.

The twin electrons and baryons are stable and can contribute to the DM density.
If the asymmetry of the twin particles and anti-particles are negligible,
the twin particles annihilate
when they freeze-out from the twin thermal bath.
While the twin particles become the asymmetric DM
if the asymmetry is non-negligible also in the twin sector.
Thus the abundance of the twin fermions will be small
if the annihilation is large or the asymmetry is small.

The lightest SUSY particle (LSP) in the twin sector
may also be stable due to R-parity in the same way as the LSP in the MSSM.
If $m_{\mathrm{tLSP}} > m_{3/2}$,
the twin LSP (tLSP) can decay to the gravitino plus the SUSY partner of the tLSP,
or to the MSSM sparticle through the gravitino, depending on the mass spectrum.
For example, the tLSP can decay to a gravitino via the processes
 $\tnu^\prime \to \nu^\prime \psi_{3/2}$ or $\wt{B}^\prime \to \gamma^\prime \psi_{3/2}$
as long as it is kinematically allowed.
The gravitino can then decay to the LSP in the MSSM.
Here we assume that the $m_{\mathrm{tLSP}} > m_{3/2} > m_\mathrm{LSP}$.
In this case, the twin LSP should be heavier than the TeV scale,
so that the twin LSP/gravitino decay does not alter the success
of Big Bang Nucleosynthesis (BBN).

\subsection{Summary}

\begin{itemize}
\item We studied the neutrino sector in a global $SU(5)$ F theory GUT.
\item $SU(5)$ is spontaneously broken to the SM gauge symmetry via a Wilson line.
\item At low energies the model has the MSSM spectrum with a complete MSSM$^\prime$ twin sector.
\item  The right-handed neutrinos in the visible and twin sectors are identified.
\item  Assuming 3 right-handed neutrinos which get a Majorana mass at a scale of order $10^{12}$ GeV,
we analyzed the light neutrino spectrum.   
\item Three predominantly sterile neutrinos get mass via the See-Saw mechanism at tree level.
\item  The other three predominantly active neutrinos obtain mass via radiative corrections.
\item  We fit the light neutrino masses to data.
\item  Questions of cosmology are saved for the future.
\end{itemize}

\section*{Acknowledgments}
This work would not have been possible without the discussions with Herb Clemens.
S.R.~received partial support for this work from DOE/ DE-SC0011726.

{\small
\bibliographystyle{JHEP}
\bibliography{refs_twin}
}

\end{document}